\documentclass[11pt]{article}
\usepackage[utf8]{inputenc}
\usepackage[T1]{fontenc}
\usepackage[margin = 1in]{geometry}
\usepackage[dvipsnames]{xcolor}
\usepackage{graphicx}
\usepackage{grffile}
\usepackage{longtable}
\usepackage{wrapfig}
\usepackage{rotating}
\usepackage[normalem]{ulem}
\usepackage{amsmath}
\usepackage{textcomp}
\usepackage{amssymb}
\usepackage{capt-of}
\usepackage{hyperref}
\usepackage{booktabs}
\usepackage{setspace}
\usepackage{mdframed}
\usepackage{setspace}
\usepackage{natbib}
\usepackage{fancyhdr}

\usepackage{float}

\author{Zachary Barnett-Howell\thanks{Yale University and Y-RISE. Contact: zachary.barnett-howell@yale.edu} \and Ahmed Mushfiq Mobarak\thanks{Yale University, Y-RISE, NBER, CEPR and IGC. Contact: ahmed.mobarak@yale.edu} }
\title{The Benefits and Costs of Social Distancing in Rich and Poor Countries}

\begin{document}

\maketitle

\begin{abstract}
    
Social distancing is the primary policy prescription for combating the COVID-19 pandemic, and has been widely adopted in Europe and North America. We estimate the value of disease avoidance using an epidemiological model that projects the spread of COVID-19 across rich and poor countries. Social distancing measures that ``flatten the curve" of the disease to bring demand within the capacity of healthcare systems are predicted to save many lives in high-income countries, such that practically any economic cost is worth bearing. These social distancing policies are estimated to be less effective in poor countries with younger populations less susceptible to COVID-19, and more limited healthcare systems, which were overwhelmed before the pandemic. Moreover, social distancing lowers disease risk by limiting people's economic opportunities. Poorer people are less willing to make those economic sacrifices. They place relatively greater value on their livelihood concerns compared to contracting COVID-19. Not only are the epidemiological and economic benefits of social distancing much smaller in poorer countries, such policies may exact a heavy toll on the poorest and most vulnerable. Workers in the informal sector lack the resources and social protections to isolate themselves and sacrifice economic opportunities until the virus passes. By limiting their ability to earn a living, social distancing can lead to an increase in hunger, deprivation, and related mortality and morbidity. Rather than a blanket adoption of social distancing measures, we advocate for the exploration of alternative harm-reduction strategies, including universal mask adoption and increased hygiene measures.

\end{abstract}

\doublespacing

The COVID-19 pandemic has generated furious debate about what public health measures might prove most effective at containing the disease and minimizing mortality rates. Without a vaccine for the novel coronavirus, governments have implemented social distancing and quaratine measures in China, Europe, and the United States designed to ``flatten the curve.'' The goal is to slow the spread of the virus to reduce immediate pressure on the healthcare system, minimizing transmission rates so that targeted testing and tracking measures can be effective in combating the disease. A parallel conversation has emerged about the economic devastation caused by such measures, especially as the virus reaches low-income countries. As social distancing becomes the universal strategy against COVID-19, a question emerges: are shuttering the economy for weeks or months and mass unemployment reasonable costs to pay? 

The answer for the United States and other rich countries appears to be yes. By assigning an economic value to the risk of COVID-19 mortality predicted in epidemiological models, it becomes clear that the cost of \textit{not} intervening in rich countries would be greater than the deepest economic recession. In other words, according to any reasonable benefit-cost metric, social distancing interventions and aggressive suppression are overwhelmingly justified.

The purpose of this note is to quantitatively explore whether similar mitigation and suppression strategies are equally valuable in low and middle income countries. There are a number of demographic and infrastructural reasons why the costs of COVID-19 and the benefits to social distancing may vary with the income level of a country. There is a greater prevalence of comorbidities and endemic disease in poor countries, such as malnutrition and tuberculosis, which may negatively interact with COVID-19 infection. The healthcare systems in poor countries have few hospital beds and ventilators per capita, and are predicted to be unable to absorb a rapid influx of COVID-19 patients. This means, however, that flattening the curve of the disease to fall within the capacity of the healthcare system may not be feasible, no matter the extent of the lockdown or mitigation efforts employed. Many more workers in poor countries are self-employed or in the informal sector and depend on daily wages to feed their families. In the absence of strong social protection and insurance, the cost imposed by social (and economic) distancing may be large in terms of immediate deprivation and hunger. As a result compliance rates with lock-down orders or social distancing guidelines may be lower in countries with weaker enforcement capacity. Most importantly however, the high average age in rich countries drives the predicted high mortality from the unmitigated spread of COVID-19, while the younger population in low-income yields a much lower predicted mortality risk.

To determine the relative value of suppression strategies in rich versus poor countries, we embed estimates of the country-specific costs of mortality developed by \citet{viscusi2017} into the influential epidemiological model developed by the the Imperial College London COVID-19 Response Team that predicts mortality from the spread of the virus \citep{ferguson2020,walker2020}. \citet{greenstone2020} adapt this model to assign an economic value to COVID-19 mortality in the United States. They predict that social distancing measures will save 1.76 million lives (both directly, and indirectly by reducing hospital overcrowding), with a total welfare value of 7.9 trillion dollars. Widespread social distancing and stay-at-home orders may create economic hardship in the United States, but this leaves no room for debate about the value of this public health intervention. We conduct a similar exercise for all rich and poor countries to explore whether such a policy prescription applies uniformly, or whether more nuanced thinking, analysis, and strategizing is required in the case of low-income countries.

\section{Methods and Results}

\subsection{Mortality}

\citet{walker2020} predicts COVID-19 from all countries across the five policy scenarios using a base R0 value of 3.0. This includes predictions for an unmitigated epidemic, where the government takes no action to limit the spread of COVID, and social distancing measures. Walker et al. predict the mortality impact of a uniform 45\% reduction in interpersonal contact rates within a country (social distancing), and social distancing with enhanced targeting of individuals over the age of 70 who further reduce their contact rates by 60\% (social distancing+). The authors also consider suppression strategies, defined by more a intensive social distancing involve a 75\% reduction in interpersonal contact rates, triggered late into the spread of COVID-19, at 1.6 deaths per 100,000 people per week, or relatively early, at 0.2 deaths per 100,000 people per week.
 
\begin{figure}[htbp!]
  \centering
  \caption{What is mortality risk from COVID-19?}
  \includegraphics[width = .48\textwidth, keepaspectratio]{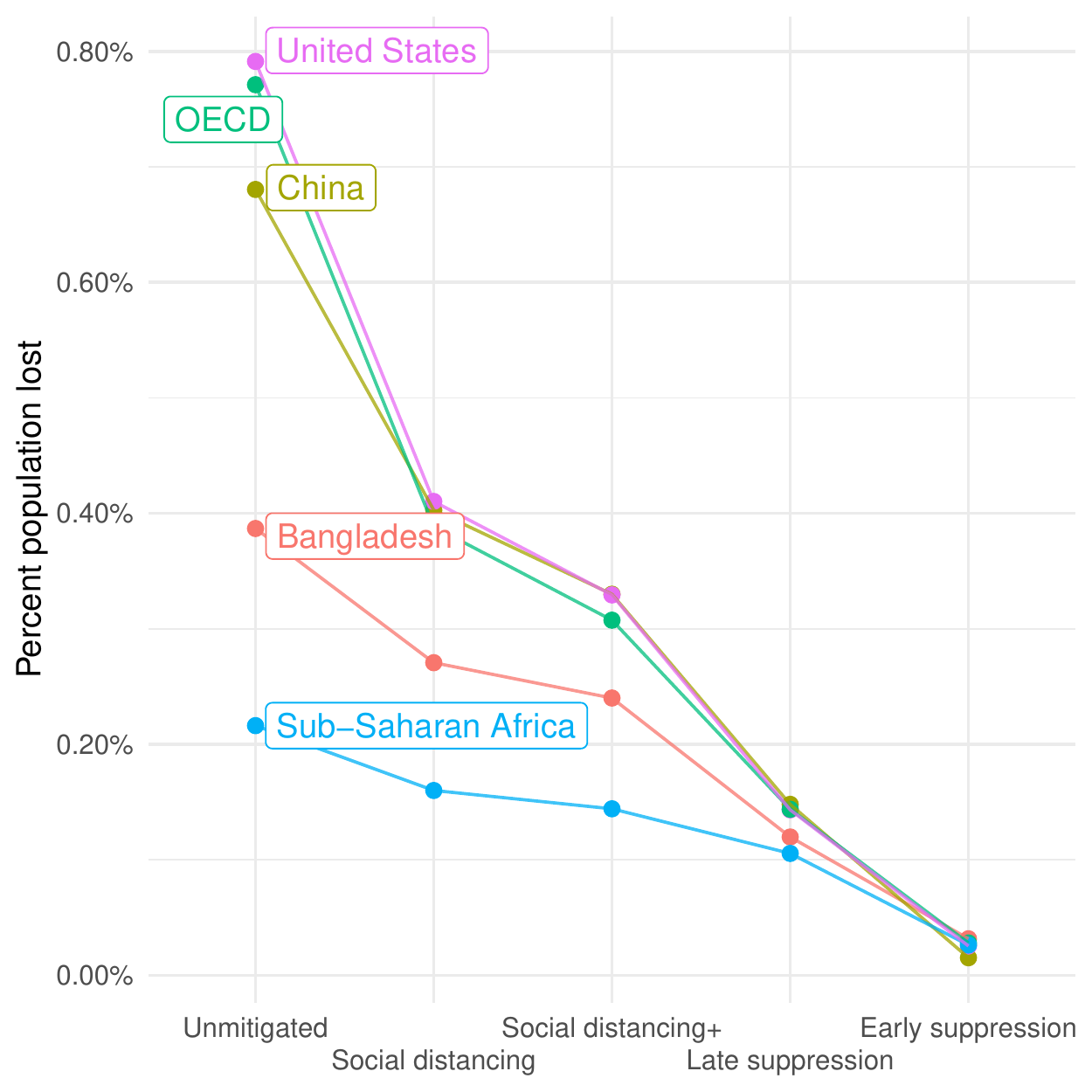}
    \includegraphics[width = .48\textwidth, keepaspectratio]{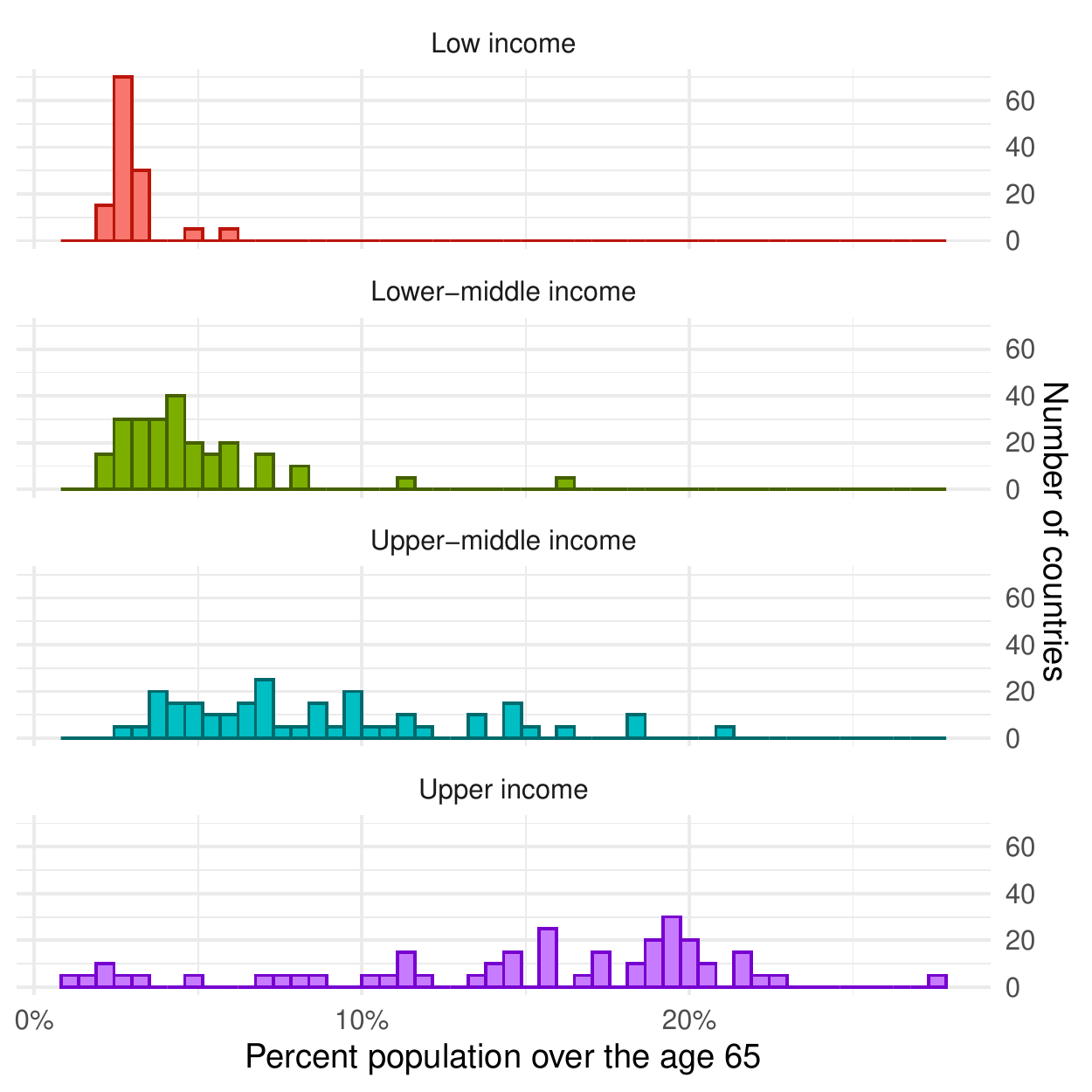}
  \label{fig:perc-mortality}
\end{figure}

Figure \ref{fig:perc-mortality} shows the predicted average mortality from COVID-19 for a set of countries and regions as well as distribution of the percentage of the population above the age of 65 by World Bank income classification. Richer regions, where the population skews older, risk losing more lives in an unmitigated outbreak. Predicted population level mortality rates are just below 0.8\% in the United States as in other OECD economies.\footnote{
    The global mortality predicted from the unmitigated spread of COVID-19 is over 135 million. For context, the H1N1 Spanish influenza of 1918 is estimated to have killed between 50 to 100 million people, somewhere between .95\% and 5.4\% of the world population at the time. See: \citet{taubenberger2006,johnson2002}.}
Countries and regions with younger populations, such as Bangladesh and Sub-Saharan Africa, face much lower risk, with the unmitigated spread of COVID-19 leading to predicted mortality rates of 0.39\% and 0.21\%, respectively. This is \emph{despite} the comparatively poor health system capacity in poorer countries (proxied by hospital beds per capita) incorporated into the Imperial College model. The right panel of the figure explains why. The proportion of the population that is elderly varies greatly between low-income (3\%) and high-income countries (17.4\%).

\subsection{Assumptions Driving the Mortality Predictions}

The prediction of significantly lower incidence of COVID-19 deaths in poor countries is primarily based on the younger average age of their population. The model accounts for the fact that poor countries have fewer hospital beds and lower ICU capacity, and will be entirely unable to meet peak demand. The lower marginal benefits of implementing suppression policies in poor countries arises from the fact that by the time suppression is triggered, the model predicts that COVID-19 will have already spread significantly, overwhelming countries with low healthcare capacity. Older people in low-income countries are also more likely to become infected by COVID-19 as they have higher contact with other individuals inside and outside the household, but the large demographic differences between rich and poor countries outweighs this factor.

The model, however, does not presently account for the higher burden of infectious diseases and chronic illness in low-income countries, particularly in children, basing its estimate of healthcare demand and overall mortality on data from China. This could lead to an under-estimate of mortality in low-income countries \citep{walker2020}. On the other hand, the model presumes equally effective implementation of mitigation or suppression policies in poor and rich countries. Recent experience in India with the large and slow exodus of migrant workers from cities following lock-down suggests that suppression policies imperfectly implemented in low-capacity settings may have counter-productive effects on containing COVID-19.

\subsection{Differences in the Economic Value of Interventions in Rich and Poor Countries}

The COVID-19 mitigation strategies considered in our model are all based on reducing contact rates. However, lower contact comes at the cost of reduced economic activity and lower earnings. We measure the economic value of avoided mortality from mitigation policies in each country using \citet{viscusi2017}'s country-specific value of statistical life (VSL) estimates. The VSL is based on how people trade off the risk of harm and economic reward. Individuals face mortality risks, whether from disease or as a result of their occupation, and individuals may accept these risks willingly when appropriately compensated. A recent example is the strike by Instacart workers in the United States during this pandemic, who are demanding an additional \$5 in hazard pay per order as compensation for their increased exposure to the disease \citep{wapo2020}. The VSL adds up these probabilistic risks to estimate the monetary value that people in a given country assign to saving one of these \textit{statistical} lives. The VSL is not a measure of the economic productivity that a person provides, but rather how society as a whole assigns value to saving a life.

\begin{figure}[htbp!]
  \centering
  \caption{What is the total VSL lost for each country?}
  \includegraphics[width = \textwidth, height = 4in, keepaspectratio]{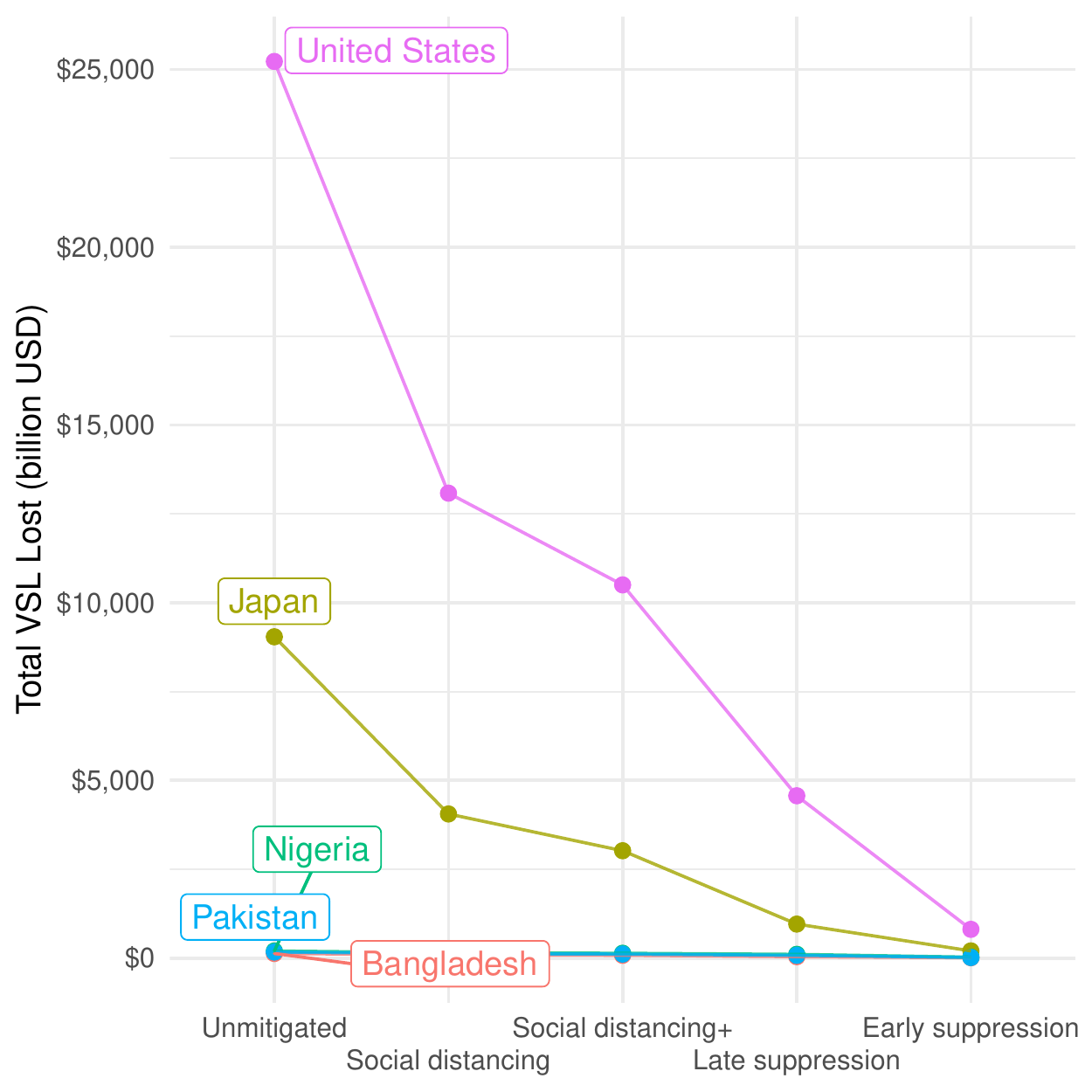}
  \label{fig:vsl-levels}
\end{figure}

Figure \ref{fig:vsl-levels} displays the estimated dollar value of total losses from deaths under each intervention scenario when the \citet{viscusi2017} VSL estimates are embedded in the Imperial College mortality predictions. The cost of leaving COVID-19 uncontrolled in the United States is unambiguously large. This is due to higher predicted mortality rates in the United States relative to other countries and the higher base VSL. In comparison to U.S. losses, the dollar costs of uncontrolled COVID-19 in large countries such as Pakistan or Nigeria look minuscule. The more relevant question for any country-specific policy is the total cost of COVID-19 mortality under each scenario relative to that country's own GDP. 

\begin{figure}[htbp!]
  \centering
  \caption{What is the relative VSL lost for each country?}
  \includegraphics[width = \textwidth, height=4.5in, keepaspectratio]{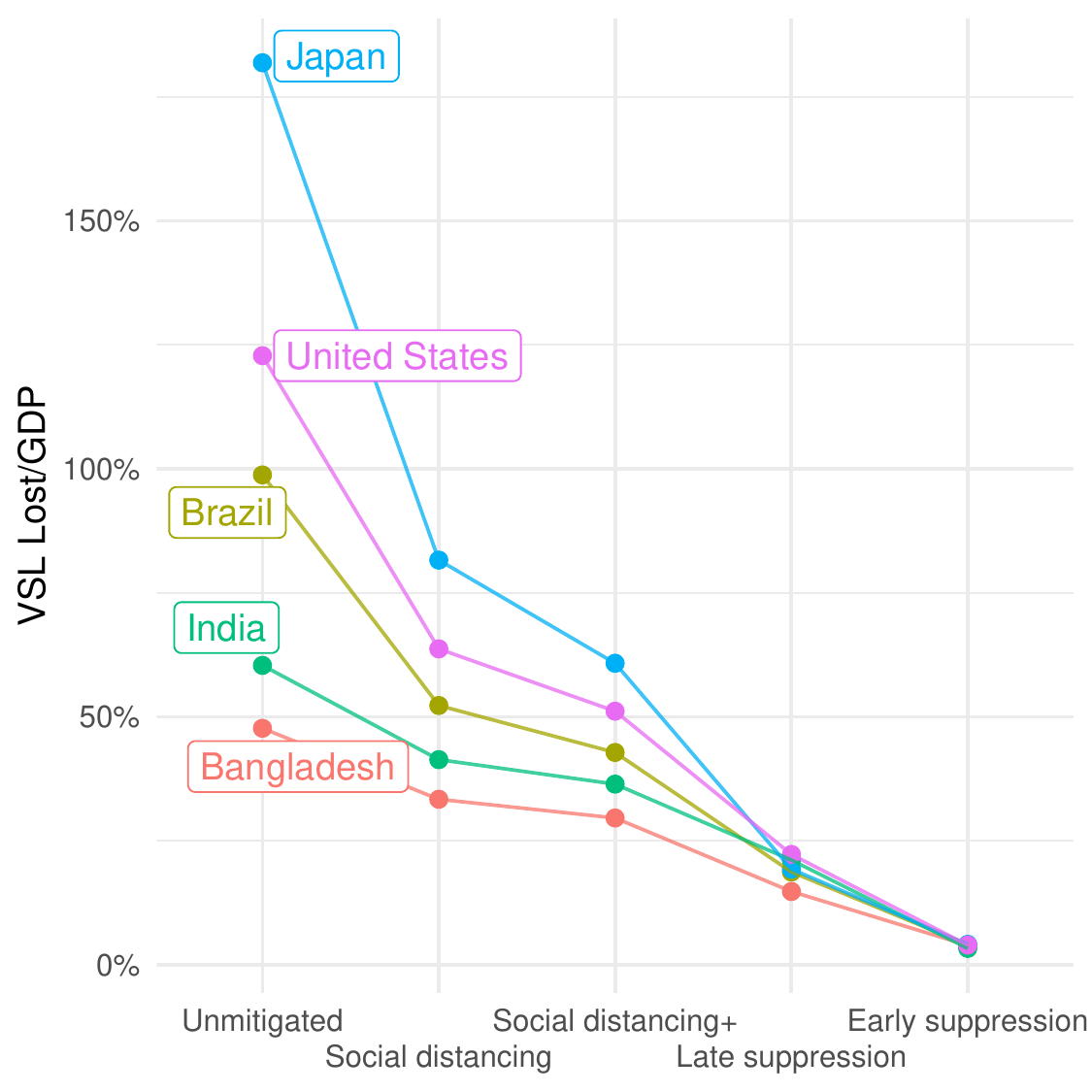}
  \includegraphics[width = \textwidth, height=4.5in, keepaspectratio]{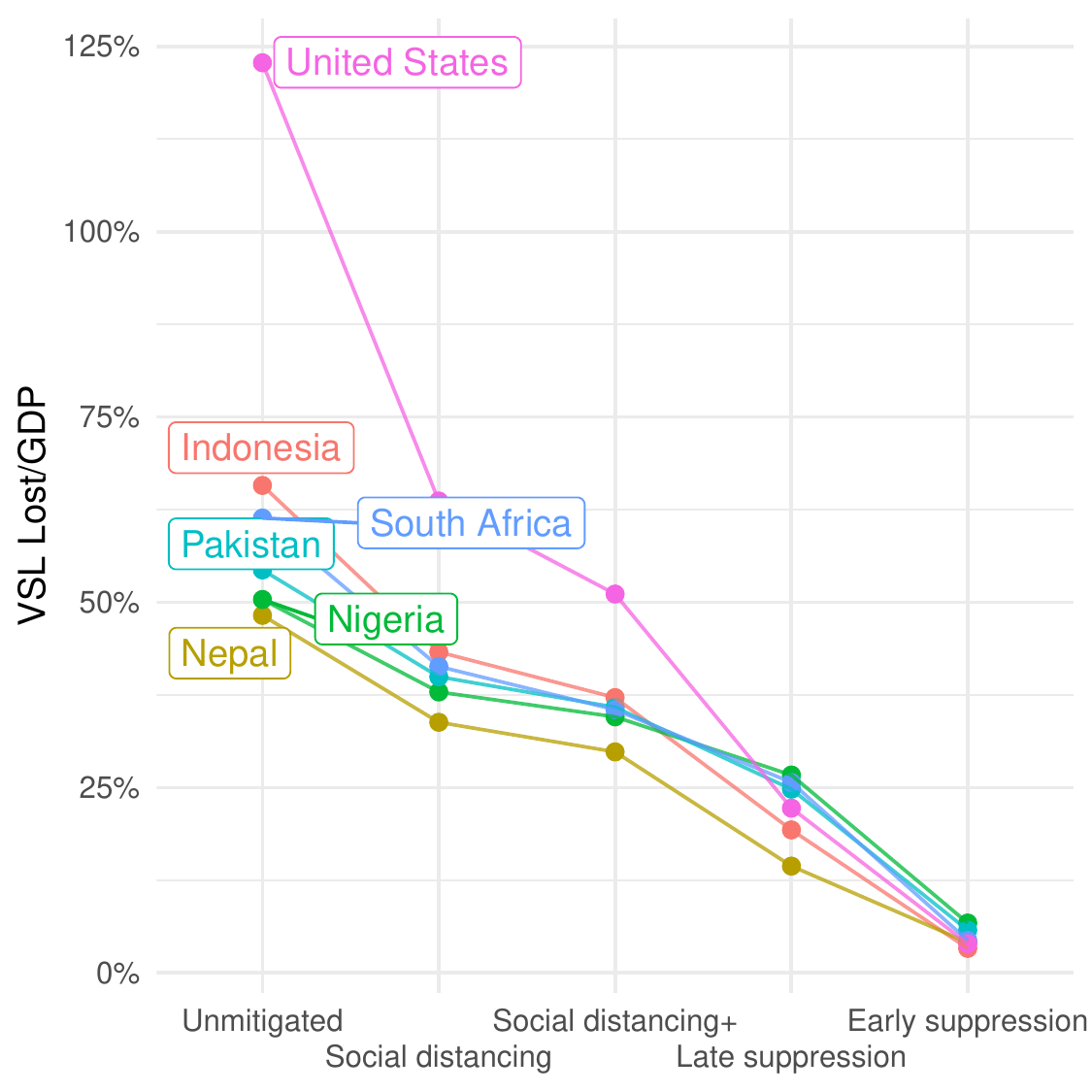}
  \label{fig:vsl-gdp}
\end{figure}

Although the value of intervention is more comparable by this metric, without mitigation efforts COVID-19 still imposes a large welfare cost\textemdash above 130\% of GDP in rich countries like the United States and Japan. In contrast, in the unmitigated scenario the losses in India, Bangladesh, Pakistan, Nigeria, Nepal are about 50-60\% of their own annual GDP. The second important lesson from Figure \ref{fig:vsl-gdp} is that moving from a policy of doing nothing to imposing social distancing yields a very large welfare improvement in rich countries: Equivalent to 59\% of U.S. GDP. However, the same policy of social distancing increases estimated welfare in Bangladesh by only 14\% of its own lower GDP. Imposing more suppressive policies in the United States yield additional increases in welfare equivalent to 42\% of its GDP, while moving from social distancing to suppression in Bangladesh only increases welfare by an additional 19\% of its own (lower) GDP. The relative gains of more stringent policy measures against COVID-19 are shown in Table \ref{tab:marginal-value-intervention} for countries at varying income levels. 

Underlying the relatively modest benefit estimates from mitigation and suppression policies in poorer countries are three critical factors. First, in poor countries there are fewer old people to who can benefit from targeted distancing. Further, the elderly often reside with younger family members, so contact rates can only be reduced within limits. Second, the relatively low hospital and ICU capacity at baseline in poorer countries means that flattening the mortality curve is unlikely to prevent hospitals from being overwhelmed. Third, the opportunity cost of social distancing is larger in poorer countries, and the VSL is therefore lower. Simply put, rich people can more easily meet their basic needs while social distancing, while a poor person may need to prioritize income-generating opportunities to put food on their family's table.

\subsubsection{Limitations of Using VSL to Value Mortality Reduction}

One concern with using the VSL to estimate welfare loss from COVID-19, is that income levels play a significant role in determining individual willingness to accept compensation for increased risks. That people in poor countries accept greater risks for lower compensation, leading to a lower estimate of their VSL, may be due to necessity rather than choice. Moreover, the VSL is estimated using very small changes in the relative risk of dying, on the order of 1:10,000. The different COVID-19 mitigation scenarios under consideration shift estimated mortality rates more drastically: two to three orders of magnitude larger. For example, moving from no mitigation to social distancing in Bangladesh reduces average risk by 1:1,000, and moving from social distancing to late suppression reduces average risk by a similar amount. It is unclear whether it is appropriate to extrapolate the  VSL, estimated from small increases in relative risk, to the much larger risks from COVID-19. Our estimates of the value of each strategy are likely to underestimate the welfare losses to COVID-19 in countries with older populations, where the change in relative risk under strategy is the largest.

\section{Differences in the Costs of Interventions in Rich and Poor Countries}

\begin{figure}[htbp!]
\centering
\label{fig:vsl-employment-income}
\caption{Estimated Value of COVID-19 Intervention by Income Group}
\includegraphics[width = .48\textwidth, keepaspectratio]{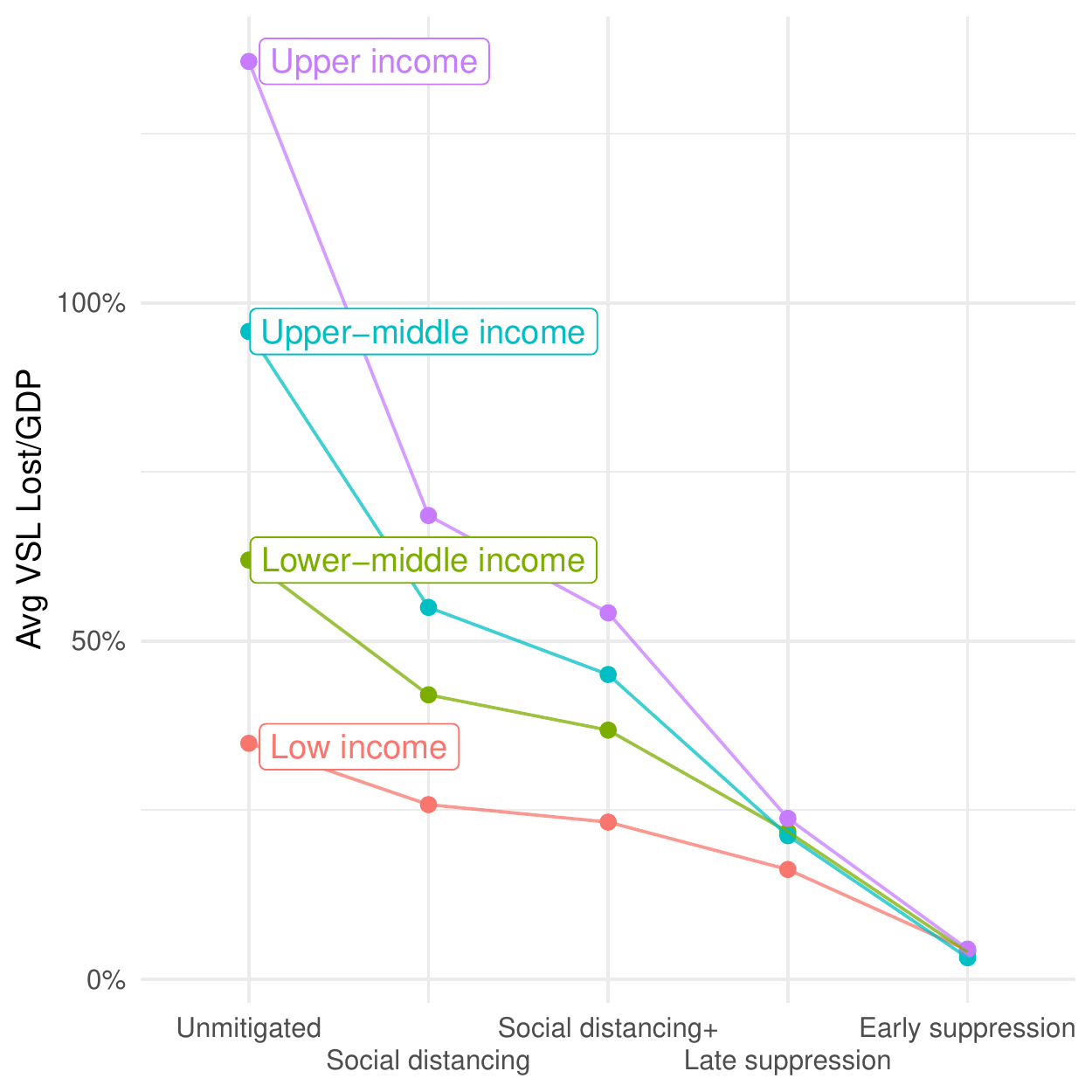}
\includegraphics[width = .48\textwidth, keepaspectratio]{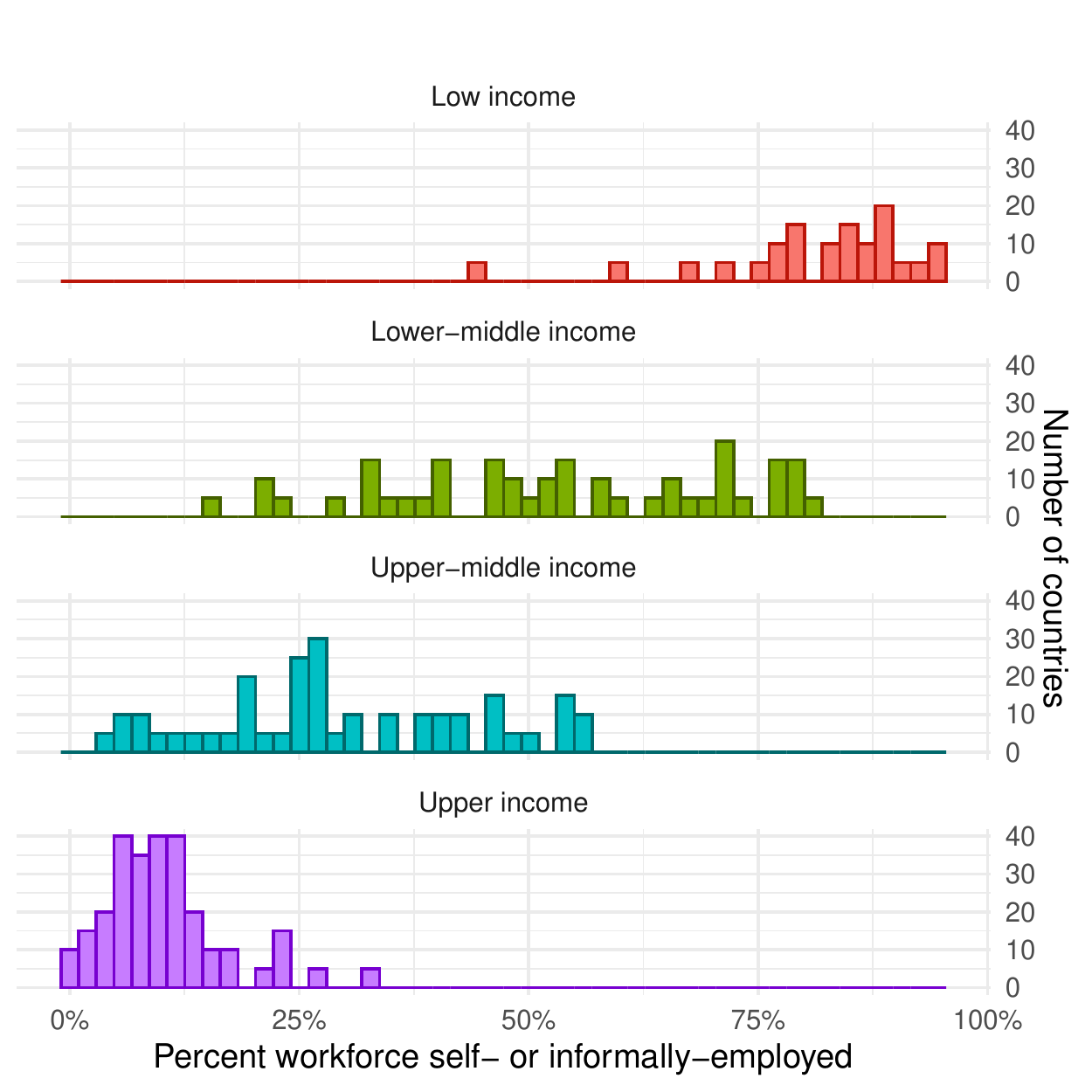}
\end{figure}

Beyond the much smaller benefits of COVID-19 mitigation in poorer countries, workers in such countries are also more vulnerable to the disruption of the economy. They are more likely to rely on a daily cash wage, their work is hands-on and cannot be done while social distancing. Figure \ref{fig:vsl-employment-income} shows the distribution of the percentage of workers either self- or informally-employed. Such workers do not always appear in government and bureaucratic records. So even if a social insurance policy were implemented in these countries, it is uncertain how quickly such people could be located, if at all, to deliver relief benefits to them.

The social distancing and suppression interventions pioneered in Wuhan, China, and now in place throughout Europe and parts of the United States, rely on government support systems. Many workers throughout Europe still receive their salaries, and U.S.\ taxpayers will receive a stimulus check. By contrast, efforts by the Indian government to impose a lock-down appear to have had significant negative consequences for the most vulnerable members of its population. Interviews with workers from the informal sector tell a story of impending poverty, evictions, and hunger, as their incomes and work opportunities have been curtailed. Migrant laborers in India's largest cities, now without access to employment, are without food or shelter. Many are in the process of literally walking back to their homes, with deaths along the journey already being reported.\footnote{
  See: \citet{abihabib2020b,abihabib2020a,bbc2020,tewari2020}. \citet{abihabib2020a} quote one migrant laborer saying: ``You fear the disease, living on the streets. But I fear hunger more, not corona.'' Another migrant construction worker is quoted saying ``I earn 600 rupees every day and I have five people to feed. We will run out of food in a few days. I know the risk of coronavirus, but I can't see my children hungry" \citep{bbc2020}.}
The mortality consequences to macroeconomic economic shocks are not straightforward, but cross-country results find that a 1\% decrease in GDP can lead to an increase in infant mortality between 0.24 and 0.40 per 1,000 children born \citep{baird2011}.\footnote{
	See: \citet{paxson2005,cutler2002,bhalotra2010} for country-level studies on the effect of economic shocks on health outcomes.}

\section{Policy Discussion}

The COVID-19 pandemic represents a serious threat in every country. A policy response is necessary, the benefits to each policy must be carefully weighed against the economic cost and risks imposed on that society. The most widely-cited model of COVID-19 transmission and mortality shows that we should expect fewer deaths in poor countries, and that social distancing policies in these countries produce smaller benefits. Much of this result is based on differences in the age distribution across countries, because our present understanding is that COVID-19 mortality risk increases dramatically with age. It is uncertain whether this relationship will remain robust in poorer countries where younger people have higher rates of chronic illness and endemic disease. Yet even permitting an overestimate of deaths in rich countries and an underestimate in poor countries, the differences in imputed welfare benefit remain vast. Given the deeper concerns about the risks that economic shutdowns pose on the most vulnerable members of low-income societies \citep{saleh2020}, it remains unclear whether the value of mitigation and suppression policies in poor countries outweighs the uncertain economic costs. 

We know that workers in low-income countries are younger and likely less susceptible to COVID-19. We know that workers are also more vulnerable to economic disruption, and may be unable to adhere to lockdown orders. Various government and non-governmental organizations are currently playing an important role to avert outright starvation during the pandemic by providing free meals, food supplies, and fuel to poor households. Supply chains within countries have been disrupted by lockdown measures, making it increasingly difficult to deliver food \citep{purohit2020}. \citet{ray2020} suggest permitting people under the age of 40 to work during lockdown as a way of mitigating the economic costs to COVID-19 suppression. Indeed, the recent example of India demonstrates our concern about the capacity of states to enforce suppression strategies, and where imperfect compliance may lead to an increase in transmission to other vulnerable populations \citep{scroll2020}. \citet{ravallion2020} highlights the tradeoff inherent to COVID-19 mitigation strategies between the risks of the disease, and the deprivation and hunger that will result from prolonged economic disruption. 

The social distancing policies implemented in European countries and the United States may well be entirely applicable to other parts of the world. Nevertheless, a serious assessment is urgently required to determine what other measures could effectively preserve lives and aggregate welfare. Once the source code for the Imperial College model is made available, social scientists can explore the sensitivity of benefit estimates to changes in assumptions about compliance with distancing guidelines, enforcement capacity, and other behavioral adjustments. We should explore quantitatively the benefits of alternative policies to social distancing, including harm-reduction measures that may allow people in low income countries to minimize their risk from COVID-19 while preserving their ability to put food on the table:

\begin{itemize}
    \item Increasing access to clean water and handwashing are likely to provide significant mortality risk reduction.\footnote{\citet{glassman2020}}
    \item Universal mask wearing when outside, even with home-made cloth masks.\footnote{\citet{abaluck2020}}
    \item Targeted social isolation of the elderly and other at-risk groups will save lives while permitting individuals with lower risk profiles to continue to work.\footnote{\citet{lshtm2020, favas2020}}
    \item If widespread social distancing is pursued, then efforts must be made to get food, fuel, and cash into the hands of the people most at risk of hunger and deprivation.
\end{itemize}

\clearpage

\begin{table}
\caption{\label{tab:marginal-value-intervention}Marginal Value of COVID-19 Interventions Relative to Unmitigated Scenario (total VSL/GDP)}
\centering
\begin{tabular}[t]{lrrrrrrrr}
\toprule
\multicolumn{1}{c}{ } & \multicolumn{3}{c}{Upper Income} & \multicolumn{2}{c}{Upper-Middle Income} & \multicolumn{3}{c}{Lower-Middle Income} \\
\cmidrule(l{3pt}r{3pt}){2-4} \cmidrule(l{3pt}r{3pt}){5-6} \cmidrule(l{3pt}r{3pt}){7-9}
Strategy & Japan & UK & US & Brazil & Indonesia & S. Africa & Bangladesh & Nigeria\\
\midrule
Unmitigated & -- & -- & -- & -- & -- & -- & -- & --\\
Social distancing & 103 & 81 & 62 & 42 & 23 & 21 & 16 & 13\\
Social distancing+targeting & 21 & 17 & 13 & 9 & 6 & 6 & 4 & 4\\
Late suppression & 42 & 33 & 30 & 22 & 18 & 10 & 16 & 8\\
Early suppression & 16 & 28 & 19 & 13 & 16 & 22 & 12 & 21\\
\bottomrule
\end{tabular}
\end{table}

\clearpage
\begin{singlespacing}

\bibliography{covid19.bib}

@techreport{greenstone2020,
  author      = {Greenstone, Michael and Nigam, Vishan},
  title       = {Does Social Distancing Matter?},
  year        = {2020},
  month       = {3},
  day         = {30},
  type        = {Sociolinguistic Working Paper},
  number      = {2020-26},
  institution = {Becker Friedman Institute for Economics},
  doi = {http://dx.doi.org/10.2139/ssrn.3561244}
  }

@article{bhalotra2010,
  title={Fatal fluctuations? Cyclicality in infant mortality in India},
  author={Bhalotra, Sonia},
  journal={Journal of Development Economics},
  volume={93},
  number={1},
  pages={7--19},
  year={2010},
  publisher={Elsevier}
}

@article{paxson2005,
  title={Child health and economic crisis in Peru},
  author={Paxson, Christina and Schady, Norbert},
  journal={The World Bank Economic Review},
  volume={19},
  number={2},
  pages={203--223},
  year={2005},
  publisher={Oxford University Press}
}

@article{cutler2002,
  title={Financial crisis, health outcomes and ageing: Mexico in the 1980s and 1990s},
  author={Cutler, David M and Knaul, Felicia and Lozano, Rafael and M{\'e}ndez, Oscar and Zurita, Beatriz},
  journal={Journal of Public Economics},
  volume={84},
  number={2},
  pages={279--303},
  year={2002},
  publisher={Elsevier}
}

@article{baird2011,
  title={Aggregate income shocks and infant mortality in the developing world},
  author={Baird, Sarah and Friedman, Jed and Schady, Norbert},
  journal={Review of Economics and Statistics},
  volume={93},
  number={3},
  pages={847--856},
  year={2011},
  publisher={MIT Press}
}

@article{wapo2020,
    author  = {Olson, Alexandra, and Anderson, Mae},
    title = {Some Instacart, Amazon workers strike as jobs get riskier},
    year = {2020},
    month = {3},
    day = {30},
    date = {2020-03-30},
    journaltitle = {Washington Post},
    urldate = {2020-04-03}
    }

@misc{ray2020,
    author = {Ray, Debraj and Subramanian, S.},
    title = {India Under Lockdown},
    date = {2020-03-27},
    year = {2020},
    url = {http://debrajray.blogspot.com/2020/03/india-under-lockdown.html},
    urldate = {2020-04-04}
}

@misc{ravallion2020,
    author = {Ravallion, Martin},
    title = {On the virus and poor people in the world},
    date = {2020-04-02},
    year = {2020},
    url = {https://economicsandpoverty.com/2020/04/02/on-the-virus-and-poor-people-in-the-world/},
    urldate = {2020-04-04}
}

@misc{saleh2020,
    author = {Saleh, Asif and Cash, Richard A.},
    title = {Masks and Handwashing vs. Physical Distancing: Do We Really Have Evidence-based Answers for Policymakers in Resource-limited Settings?},
    date = {2020-04-03},
    year = {2020},
    url = {https://www.cgdev.org/blog/masks-handwashing-vs-physical-distancing-do-we-really-have-evidence-based-answers},
    url = {2020-04-04}
}

@article{scroll2020,
    author  = {Agrawal, Parul},
    title = {COVID-19 lockdown: Bihar migrants who fled cities face ostracism at home},
    year = {2020},
    month = {4},
    day = {2},
    date = {2020-04-02},
    journaltitle = {Scroll.in},
    urldate = {2020-04-03},
    url = {https://scroll.in/article/958010/covid-19-lockdown-bihar-migrants-who-fled-cities-face-ostracism-at-home}
    }

@article{bbc2020,
    author = {BBC},
    title = {India's poorest 'fear hunger may kill us before coronavirus},
    journaltitle = {BBC News},
    year = {2020},
    date = {2020-03-25},
    url = {https://www.bbc.com/news/world-asia-india-52002734},
    urldate = {2020-04-02}
}

@article{taubenberger2006,
  title={1918 Influenza: the mother of all pandemics},
  author={Taubenberger, Jeffery K and Morens, David M},
  journal={Emerging Infectious Diseases},
  volume={12},
  number={1},
  pages={15},
  year={2006},
  publisher={Centers for Disease Control and Prevention}
}

@article{abihabib2020a,
    author = {Abi-Habib, Maria and Yasir, Sameer},
    title = {India's Coronavirus Lockdown Leaves Vast Numbers Stranded and Hungry},
    journal = {New York Times},
    date = {2020-03-29},
    url = {https://www.nytimes.com/2020/03/29/world/asia/coronavirus-india-migrants.html},
    urldate = {2020-03-31}
}

@article{abihabib2020b,
    author = {Abi-Habib, Maria and Yasir, Sameer},
    title = {For India's Laborers, Coronavirus Lockdown Is an Order to Starve},
    journal = {New York Times},
    date = {2020-03-30},
    url = {https://nyti.ms/2JrIPEU},
    urldate = {2020-03-31}
}

@article{tewari2020,
    author = {Tewari, Abhimanyu},
    title = {The Delhi exodus: What we failed to learn from 1889 and 1900},
    journal = {The Indian Express},
    date = {2020-03-31},
    year = {2020},
    url = {https://indianexpress.com/article/opinion/the-delhi-exodus-what-we-failed-to-learn-from-1889-and-1900-coronavirus-india-lockdown-6340334/},
    urldate = {2020-03-31}
}

@article{viscusi2017,
  title={Income elasticities and global values of a statistical life},
  author={Viscusi, W Kip and Masterman, Clayton J},
  journal={Journal of Benefit-Cost Analysis},
  volume={8},
  number={2},
  pages={226--250},
  year={2017},
  publisher={Cambridge University Press}
}

@article{johnson2002,
  title={Updating the accounts: global mortality of the 1918-1920  ``Spanish" influenza pandemic},
  author={Johnson, Niall PAS and Mueller, Juergen},
  journal={Bulletin of the History of Medicine},
  pages={105--115},
  year={2002},
  publisher={JSTOR}
}

@unpublished{walker2020,
 title={The Global Impact of COVID-19 and Strategies for Mitigation and Suppression},
 month={3},
 date={2020-03-26},
 year={2020},
 author={Walker, Patrick GT and Whittaker, Charles and Watson et al., Oliver},
 doi = {https://doi.org/10.25561/77735}
}

@misc{lshtm2020,
    institution = {London School of Hygiene \& Tropical Medicine},
    title = {COVID-19 control in low-income settings and displaced populations: what can realistically be done?},
    date = {2020-03-20},
    url = {https://www.lshtm.ac.uk/research/centres/health-humanitarian-crises-centre/news/102976},
    urldate = {2020-04-05}
}

@unpublished{favas2020,
    title = {Guidance for the prevention of COVID-19 infections among high-risk individuals in camps and camp-like settings},
    author = {Favas, Caroline},
    institution = {London School of Hygiene \& Tropical Medicine},
    date = {2020-03-31},
}

@misc{glassman2020,
    author = {Glassman, Amanda and Chalkidou, Kalipso and Sullivan, Richard},
    title = {Does One Size Fit All? Realistic Alternatives for COVID-19 Response in Low-Income Countries},
    date = {2020-04-02},
    url = {https://www.cgdev.org/blog/does-one-size-fit-all-realistic-alternatives-covid-19-response-low-income-countries},
    urldate = {2020-04-05}
}

@article{purohit2020,
    author = {Purohit, Kunal},
    title = {India COVID-19 lockdown means no food or work for rural poor},
    date = {2020-04-03},
    year = {2020},
    journal = {Al Jazeera},
    url = {https://www.aljazeera.com/news/2020/04/india-covid-19-lockdown-means-food-work-rural-poor-200402052048439.html},
    urldate = {2020-04-05}
}

@unpublished{abaluck2020,
    author = {Abaluck, Jason and Chevalier, Judith A. and Christakis, Nicholas A. and Forman, Howard Paul and Kaplan, Edward H. and Ko, Albert and Vermund, Sten H.},
    title = {The Case for Universal Cloth Mask Adoption and Policies to Increase Supply of Medical Masks for Health Workers},
    year = {2020},
    month = {4},
    date = {2020-04-02}
}

@unpublished{ferguson2020,
  author = {Ferguson, NM and Laydon, D and Nedjati-Gilani et al., G},
  title = {Impact of non-pharmaceutical interventions (NPIs) to reduce COVID-19 mortality and healthcare demand},
  year = {2020},
  date = {2020-03-16},
  doi = {https://doi.org/10.25561/77482}
}

\end{singlespacing}

\end{document}